\documentclass[pre,twocolumn]{revtex4-1}
\usepackage{amsmath}
\usepackage{graphicx}
\usepackage{color}
\usepackage[a-1b]{pdfx}

\begin{document}
\title{Machine-learning-derived protocols for information-based work extraction from active particles}
\author{Grzegorz Szamel}
\email{grzegorz.szamel@colostate.edu}
\affiliation{Department of Chemistry,
  Colorado State University, Fort Collins, CO 80523}

\begin{abstract}
  We propose and analyze a process that extracts useful work from a single active particle
  maintained at constant temperature in a harmonic potential by measuring the relative sign of
  the self-propulsion and the confining force and then adjusting the stiffness of the potential.
  First, we show analytically that useful work can be extracted by stepwise changes of the
  stiffness. Then, we use a machine learning procedure to find time-dependent stiffness change
  protocols. We find that these protocols involve discontinuous initial changes of the stiffness 
  opposite to the expected direction, which resemble initial jumps analytically found by Garcia-Millan \textit{et al.}
  [Phys. Rev. Lett. \textbf{135}, 088301 (2025)] in a different information-based work extraction process.
  The learned protocols allow to extract significantly larger amounts of useful work. 
  The work extracted exceeds that allowed by the conventional second law for 
  feedback-controlled processes, which can be rationalized
  by the non-equilibrium character of the system considered.
\end{abstract}

\maketitle

Active particles \cite{Ramaswamyrev1,Bechingerrev,Ramaswamyrev2,Marchettirev2}
use energy from their environment to perform persistent motion. As a result,
systems consisting of active particles are intrinsically out of equilibrium.
This leads to many phenomena that defy our equilibrium-based
intuition, \textit{e.g.}, motility-induced phase separation \cite{CatesARCMP} and flocking \cite{TonerARCMP}.

The non-equilibrium character of systems of active particles allows one to devise new ways to
extract useful work from such systems \cite{Sokolov2010,DiLeonardo2010,DiLeonardo2017,Pietzonka2019}. 
In particular, isothermal cyclic engines have been proposed and analyzed
\cite{Martin2018,Ekeh2020,Szamel2020,Fodor2021,Speck2022}. One can also try to exploit the non-equilibrium
character of active systems by devising novel versions of the Szilard engine
that use active particles as their working elements. A Szilard engine \cite{Szilard}, sometimes referred to as
``an information engine,'' extracts useful work from a single
heat reservoir using information about the system obtained from a measurement. Szilard engines were
originally proposed and then extensively analyzed for thermal (passive) systems
\cite{ParrondoHorowitzSagawa2015}. 

The first active-particle-based Szilard engine was proposed by
Malgaretti and Stark \cite{MalgarettiStark2022}. Their dynamic Szilard engine used
active velocity to push a wall placed in front of the particle, \textit{i.e.} in the direction of its active velocity.
Malgaretti and Stark showed that the finite correlation time of the active velocity leads to extraction of useful work.

Subsequently, Cocconi \textit{et al.} \cite{Cocconi2023}
investigated protocols for optimal work extraction from an active
particle whose trajectory in space (but not self-propulsion) is constantly monitored. They showed that
useful work can be extracted from such an information engine even if the positional trajectory of the
particle is time-symmetric.

More recently, Garcia-Millan \textit{et al.} \cite{GarciaMillan2025,Schuettler2025} extended the seminal results
of Schmiedl and Seifert \cite{Schmiedl2007} on optimal protocols for driving a single
thermal (passive) particle to active particle models.
Garcia-Millan \textit{et al.} showed that an optimal protocol conditioned
upon a measurement of an active particle's self-propulsion can extract useful work from a single active particle
in a harmonic potential by changing the position of the potential's minimum. Thus, such an optimal
protocol constituted an active information engine.

\begin{figure}
  \includegraphics[width=0.95\columnwidth]{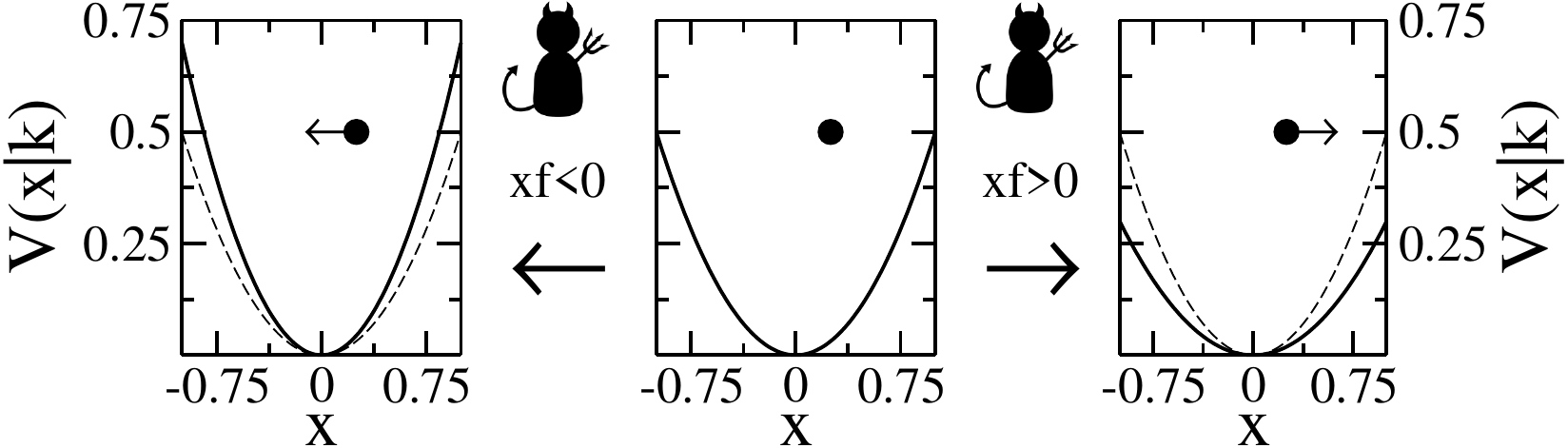}
  \caption{\label{fig0} Work extraction.
    The demon measures the relative sign between the self-propulsion and the confining force.
    When the self-propulsion (shown as the arrow attached to the particle) is aligned with the confining force,
    $xf<0$, the potential stiffness is increased. Conversely, when the self-propulsion is anti-aligned with the
    confining force, $xf>0$, the potential stiffness is decreased. At the end of the extraction process, the
  stiffness returns to its original value.}
\end{figure}

Here we propose a  way to extract useful work from a single active particle that adopts the idea of
the dynamic Szilard engine of Malgaretti and Stark to a particle in a harmonic potential.
We assume that the initial direction of the particle's self-propulsion relative to the direction of the confining
force can be measured. As shown in Fig. \ref{fig0},
if the self-propulsion points in the direction of the confining force, the stiffness
of the potential can be increased, resulting in positive work extracted from the particle. Conversely,
if the self-propulsion's direction is opposite to the confining force, the stiffness
of the potential can be decreased, again resulting in positive work extracted from the particle.
We start by assuming that the changes of the stiffness following the measurement are instantaneous:
the stiffness is decreased or increased for a certain period of time and then it returns to its initial value.
We optimize the changes of the stiffness and show explicitly that useful work can be extracted. 

Next, we use machine learning to devise time-dependent protocols that maximize the extracted work
for a given duration of the process. To this end we use the procedure proposed by
Whitelam \cite{WhitelamPRX2023} in the context of driving thermal systems
and subsequently used by Casert and Whitelam \cite{CasertWhitelam2024}
to identify protocols for driving active matter systems. The procedure uses a genetic algorithm  
to train a neural network that encodes protocols
maximizing the extracted work. We follow Casert and Whitelam and refer to these protocols as ``learned''
rather than ``optimal,'' since different training runs result in very similar but not identical protocols. 
We show that using the time-dependent learned protocols rather than stepwise decreases/increases of the
stiffness significantly increases the amount of work extracted. Surprisingly, but in agreement with the 
findings of Garcia-Millan \textit{et al.}, the learned protocols include discontinuous initial jumps of the
stiffness in the direction opposite to the one expected. 

\textit{Model} --- We consider a single self-propelled particle, in one spatial dimension,
moving in a harmonic potential $V(x|k)=kx^2/2$ with stiffness $k$. We assume overdamped dynamics. 
The equation of motion for the particle's position reads
\begin{equation}\label{eomx}
  \gamma \dot{x} = - \partial_x V(x|k) + a f + \xi,
\end{equation}
where $\gamma$ is the friction constant, $a f$ is the self-propulsion
and $\xi$ is the thermal, Gaussian white noise, $\left< \xi(t)\xi(t')\right>=2\gamma T\delta(t-t')$.
The form of the self-propulsion slightly generalizes the well-known active Ornstein-Uhlenbeck particle (AOUP)
model \cite{Szamel2014,Maggi2015,Sandford2017,AOUPreview}. We write the self-propulsion
as a product of parameter
$a$ quantifying its strength and variable $f$ that evolves according to the Ornstein-Uhlenbeck
stochastic process with independent thermal noise. Specifically, the equation
of motion for $f$ reads
\begin{equation}\label{eomf}
  \tau_p \dot{f} = -f + \eta,
\end{equation}
where $\tau_p$ is the persistence time and $\eta$ is the thermal, Gaussian white noise,
$\left< \eta(t)\eta(t')\right>=2\gamma T\delta(t-t').$ This generalized AOUP model is closer in spirit to 
the active Brownian particle (ABP) model  \cite{tenHagen2011,Fily2012}.
In the latter model the self-propulsion velocity is a product of its constant
magnitude $v_0$ and unit vector $\mathbf{e}$ that performs rotational diffusive motion
driven by thermal, Gaussian white noise. In the present model $a$ is the analog of $v_0$ and
$f$ is the analog of $\mathbf{e}$. Like for the original AOUP model \cite{Szamel2014},
the linearity of the equation of motion for $f$ significantly simplifies the analysis of the model.

In particular, for the present model, stationary state distribution, $P^{ss}(x,f)$,
for a single particle in a harmonic potential is Gaussian, $P^{ss}(x,f)\propto\exp\left[-Ax^2-Bf^2-Cxf\right]$ with
$A=(k/2T)\left[1+a^2/(1+k\tau_p/\gamma)^2\right]^{-1}$,
$B=(\tau_p/2\gamma T)\left[1+a^2/(1+k\tau_p/\gamma)\right]/\left[1+a^2/(1+k\tau_p/\gamma)^2\right]$
and $C=-\left[k\tau_pa/\gamma T(1+k\tau_p/\gamma)\right]/\left[1+a^2/(1+k\tau_p/\gamma)^2\right]$.

We use the standard stochastic thermodynamics \cite{Sekimoto1998,Seifert2012} definition 
of the work done while changing potential's stiffness $k$,
\begin{equation}\label{work1}
W = \int_0^\tau dt \, \dot{k} \, \partial_k V(x|k)  = \int_0^\tau dt \, \dot{k} \, x^2/2.
\end{equation}
We note that the same definition of work was used in the analysis
of small active particle systems \cite{Ekeh2020,Szamel2020,Fodor2021,GarciaMillan2025,Schuettler2025}.

To evaluate average work we need the time-dependent
average second moment of particle's position, $\left<x^2\right>$.
The linearity of the equations of motion (\ref{eomx}-\ref{eomf}) implies that equations of motion for
the second moments, $\left<x^2\right>$, $\left<xf\right>$ and $\left<f^2\right>$ are closed, \textit{i.e.}
they do not involve any 
higher moments \cite{Szamel2020}. This fact allows one to solve these equations either analytically (for
stepwise changes of the stiffness) or numerically (for general time dependence of the stiffness, encoded through
a neural network). 

\textit{Stepwise changes of the stiffness} --- As shown in Fig. \ref{fig0}, 
we assume that a ``demon'' measures the relative direction
of the self-propulsion and the confining force of an active AOUP in the stationary state.
In our case this amounts to measuring the sign of the product $xf$.  If
the sign is negative, the self-propulsion and the harmonic force jointly push the particle towards the
minimum of the potential. If the sign is positive, the
self-propulsion pushes against the confining harmonic force, away from the minimum of the potential.

If the self-propulsion pushes toward the minimum of the potential we make the stiffness larger,
$k\to k_1>k$ to use the finite persistence time of the self-propulsion to perform useful work.
For instantaneous, stepwise changes of the stiffness average work $\left<W\right>_-$ can be calculated
analytically. Here $\left<\ldots\right>_-$ denotes conditional averaging, with the condition that the
sign of the product $xf$ at the initial time, $t=0$, is negative. To calculate $\left<W\right>_-$ one needs to
solve equations of motions for moments $\left<x^2\right>_-$, $\left<xf\right>_-$ and $\left<f^2\right>_-$.
The average work extracted depends on the length of time $t_f$ during which the
stiffness is changed from its initial value. The long-time limit, $t_f\to\infty$, is relatively simple,
\begin{equation}\label{wm}
  \left<W\right>_- = (k_1-k)\left[ \frac{\left<x^2\right>_-(0)}{2}
    - \frac{T}{2k_1}\left(1+\frac{a^2}{1+k_1\tau_p/\gamma}\right)\right].
\end{equation}
Here $\left<x^2\right>_-(0)$ is the conditional average of $x^2$ at the initial time, $t=0$.
The form of Eq. \eqref{wm} can be easily rationalized: it is the sum of the work done on the
system during the initial increase of the stiffness to $k_1$ at $t=0$ and the work extracted from the system during
the final decrease of the stiffness back to $k$,  after the system evolved for a long time with stiffness $k_1$.

Explicit evaluation of expression \eqref{wm} shows that there is a range of $k_1$ values resulting in positive 
useful work, $- \left<W\right>_->0$. For example, for $\tau_p=1$, $a=10$, $T=1$, $\gamma=1$, and initial
stiffness $k=1$, the
largest amount of useful work $- \left<W\right>_-= 2.332$ is extracted for $k_1=1.355$.

The work extraction procedure we used resembles the dynamic Szilard engine introduced by Malgaretti
and Stark \cite{MalgarettiStark2022} in that the system is changed instantaneously for a period of time.
We emphasize that useful work is extracted due to non-trivial
correlations between the position of the active particle and its self-propulsion.
Specifically, the second term in square brackets in Eq. \eqref{wm}, which represents $\left<x^2\right>$ in the
stationary state with stiffness $k_1$, is equal to $32.08$, for stiffness $k_1=1.355$,
which is smaller than stationary state average $\left<x^2\right>$ for stiffness $k=1$, which is equal to $51$.
However, due to non-trivial correlations between the position of the active particle and its self-propulsion,
conditional average $\left<x^2\right>_-(0)$ is still smaller and equal
to $18.94$. This makes the difference is square brackets negative and useful work $-\left<W\right>_-$ positive. 

Conversely, if the self-propulsion pushes against the confining force we make the stiffness smaller,
$k\to k_2<k$ to take advantage of the direction of the self-propulsion. The expression for average work
$\left<W\right>_+$ can be obtained from Eq. \eqref{wm} by changing subscript $-\to +$ and $k_1\to k_2$.
It turns out that in this case, for a range of $k_2$ values, we can also extract useful work.
For example, for $\tau_p=1$, $a=10$, $T=1$, $\gamma=1$, and initial
stiffness $k=1$, the
largest amount of useful work $- \left<W\right>_+=0.180$ is extracted for $k_2=0.936$.

Averaging $- \left<W\right>_-$ and $- \left<W\right>_-$ over probabilities $P_-$ and $P_+$
of measuring negative and positive sign of $xf$ at $t=0$,
we get the final long-time, $t_f\to\infty$, average useful work, $-\left<W\right> = 0.724$.

We note that in our case, like in the original Szilard engine, the ``demon'' performs a one-bit measurement.
The information content of this one bit is $-P_-ln(P_-)-P_+ln(P_+)$, which for the system considered 
($\tau_p=1$, $a=10$,  $T=1$, $\gamma=1$ and 
$k=1$) gives $0.566$, which is less than $-\left<W\right>/T = 0.724$. This shows that for
our simple active information engine the conventional second law for feedback-controlled
processes \cite{ParrondoHorowitzSagawa2015}
is not applicable. We note that our analysis
does not account for the thermodynamic cost of generating the self-propulsion. More generally, 
we emphasize that the working element of our engine is a non-equilibrium system;
for such systems thermodynamics of feedback-controlled processes
is still being formulated. 

\begin{figure}
  \includegraphics[width=0.95\columnwidth]{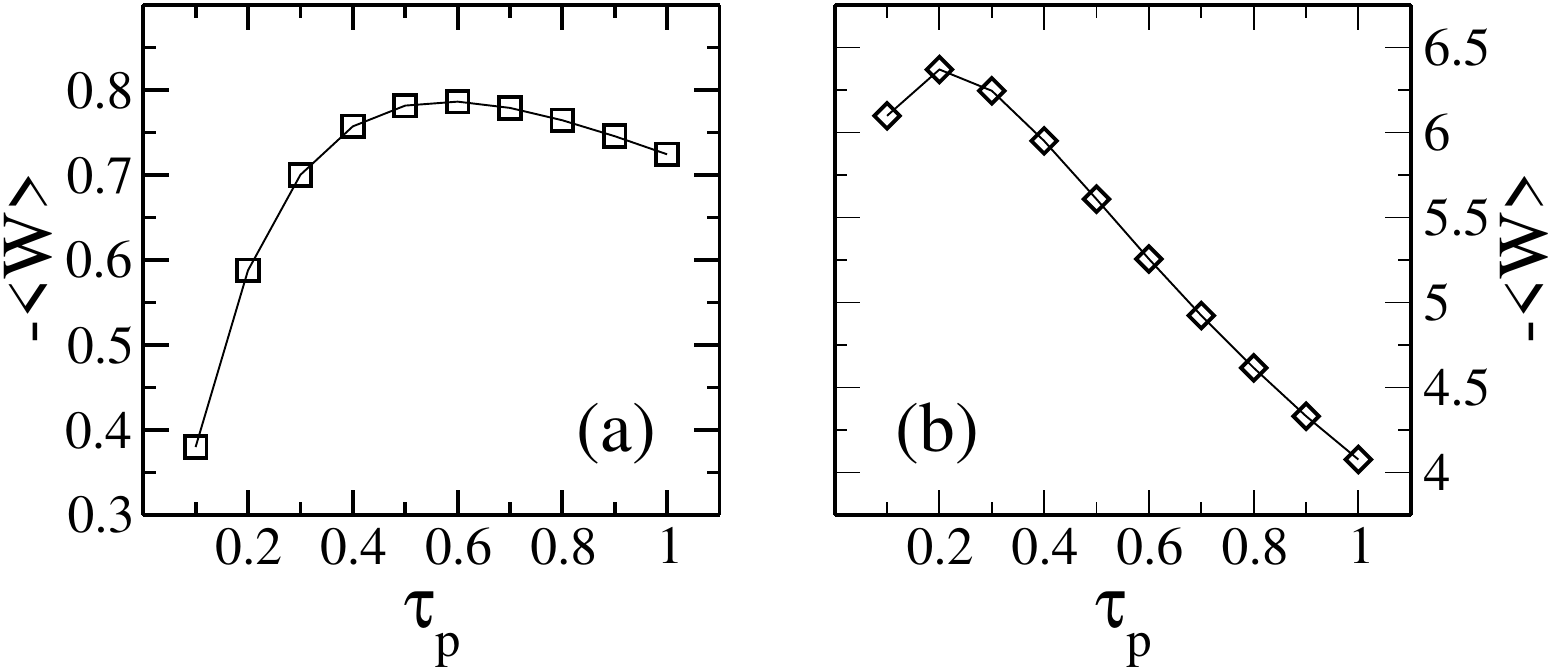}
  \caption{\label{fig1} (a) Long-time limit of average useful work,
    $-\left<W\right>$, extracted by means of stepwise changes of the
    harmonic potential stiffness, as a function of persistence time, $\tau_p$.
    (b) Average useful work extracted using learned stiffness protocols for
    a large but finite process time, $t_f=256$, as a function of persistence time, $\tau_p$.
    Using learned potential results in a significantly larger extracted work. 
    Initial and final stiffness $k=1$, self-propulsion strength $a=10$,
    temperature $T=1$ and friction constant $\gamma=1$.}
\end{figure}

Finally, in Fig. \ref{fig1}a we show the dependence of the long-time limit of the useful work $-\left<W\right>$
on the persistence time of the self-propulsion. We observe that for $a=10$ the largest amount
of work can be extracted for persistence time of about $\tau_p\approx 0.6$ (for $a=10$, $T=1$, $\gamma=1$
and $k=1$).

\textit{Machine learned stiffness protocols} --- Schmiedl and Seifert \cite{Schmiedl2007} analyzed
protocols that minimize work required to drive a single overdamped Brownian particle in a finite time,
by changing the position of the minimum and the stiffness of the harmonic potential. 
They found that optimal protocols involve discontinuous jumps of the control parameter both at the
beginning and at the end of the process. Their results for optimal protocols for changing the
position of the harmonic potential minimum were generalized to a single active particle by
Garcia-Millan \textit{et al.} \cite{GarciaMillan2025,Schuettler2025}. The latter authors showed that
discontinuous jumps of the control parameter are also present for optimal protocols controlling active
particles. Furthermore, they showed 
that the optimal protocol conditioned upon a measurement of an active particle's self-propulsion results
in negative minimal work, \textit{i.e.} in positive useful work in the surroundings. Thus they realized
an optimal work extraction procedure.

\begin{figure}
  \includegraphics[width=0.95\columnwidth]{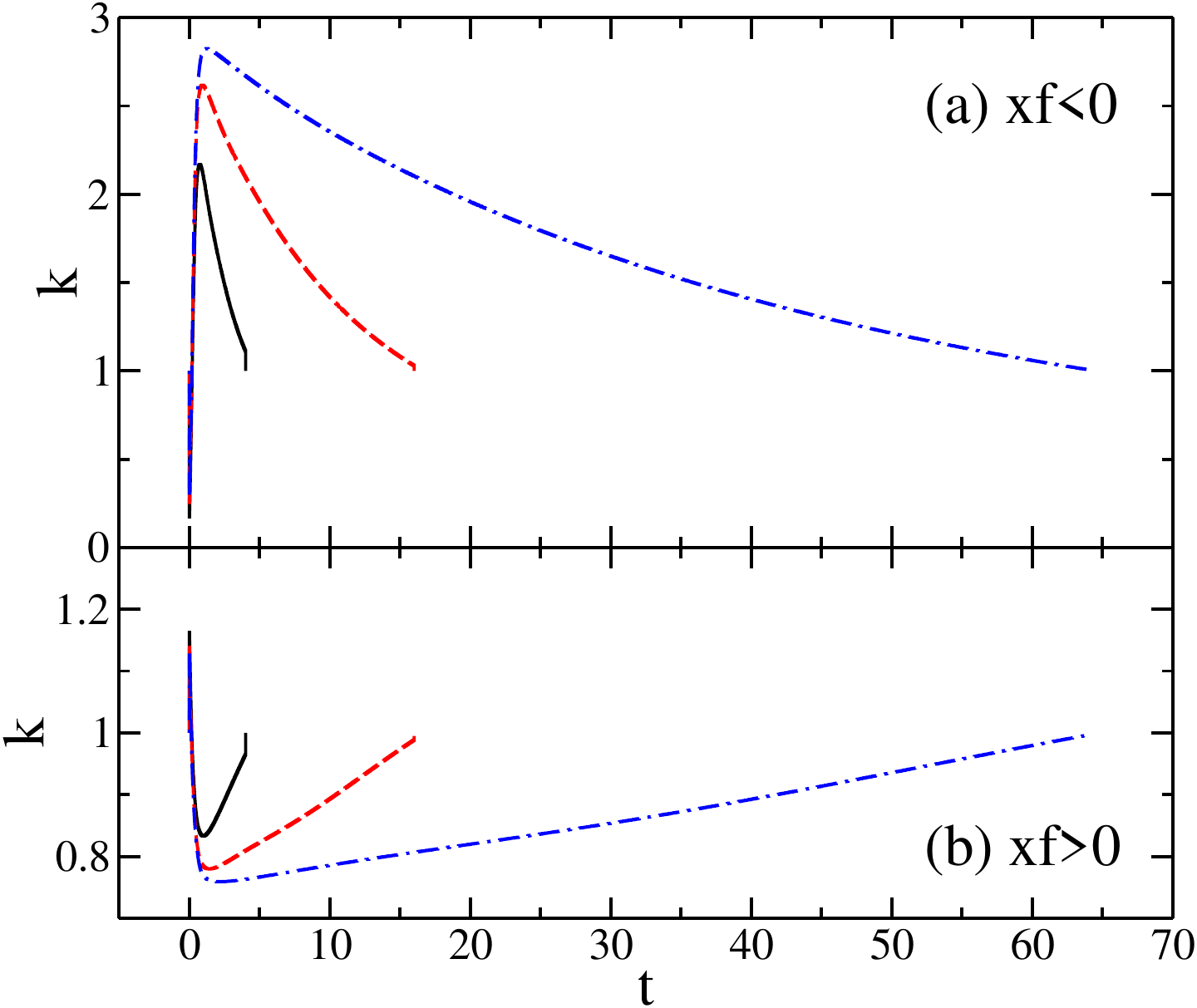}
  \caption{\label{fig2} Time-dependence of learned stiffness protocols for various lengths of the extraction
    process. Upper panel (a) presents protocols for $xf$ being negative at the initial time
    and lower panel (b) presents protocols for $xf$ being initially positive. Solid lines: $t_f=4$,
    dashed lines: $t_f=16$, dot-dashed lines: $t_f=64$. Note discontinuous jumps at the initial and final
    times. The jumps at $t=0$ are larger than jumps at $t=t_f$ and are
    in the direction opposite to the one expected, as also found in Ref. \cite{GarciaMillan2025}.
    The magnitude of the jumps decreases with increasing $t_f$. 
    Initial and final stiffness $k=1$, persistence time $\tau_p=0.4$, self-propulsion strength $a=10$, 
    temperature $T=1$ and friction constant $\gamma=1$.}
\end{figure}

Our goal is to optimize work extraction by changing the stiffness following a measurement of the relative
direction of the self-propulsion and the confining harmonic force. To this end we use a machine learning approach
originally proposed by Whitelam to find time-dependent feedback-control protocols for extracting work from
thermal systems  \cite{WhitelamPRX2023}.
This approach was subsequently used by Casert and Whitelam \cite{CasertWhitelam2024} to
find learned protocols for driving active particles with minimal heat dissipated in the process.  
Specifically, we represent the protocol, \textit{i.e.} the time dependence of the stiffness,
\textit{via} a deep neural network, which is  trained using the genetic
algorithm. The network is fully connected, with four hidden layers of width four and one additional
hidden layer of width ten. In practice, since our measurement done at the initial state of the procedure
results in a binary output, we use two independent networks for self-propulsion in the
direction of the confining force and opposing the confining force.
The input for each of the networks is time elapsed in the process and
the output is the instantaneous value of the harmonic potential stiffness.
To train each network we use a genetic algorithm \cite{WhitelamPRX2023}.
First, we generate 50 protocols by initializing
instances of the network with independent Gaussian random numbers. Next, we pick the five protocols
that result in the largest average useful work. 
We generate 49 protocols of the next generation by picking with
replacement from the set of five best protocols and adding independent random numbers to
the network parameters. The 50th protocol of the next generation is the un-modified
best protocol of the previous generation. The process is then repeated. In practice we found that
the largest useful work is extracted if the variance of the ``mutations'' is decreased during the
training process (some additional technical details are provided in the appendix).
As usual, we cannot guarantee that the final useful work is optimal. While the final
useful work is well reproducible between different learning runs, the final protocols show some differences and 
for this reason we refer to them as learned protocols rather than optimal protocols.

\begin{figure}
  \includegraphics[width=0.95\columnwidth]{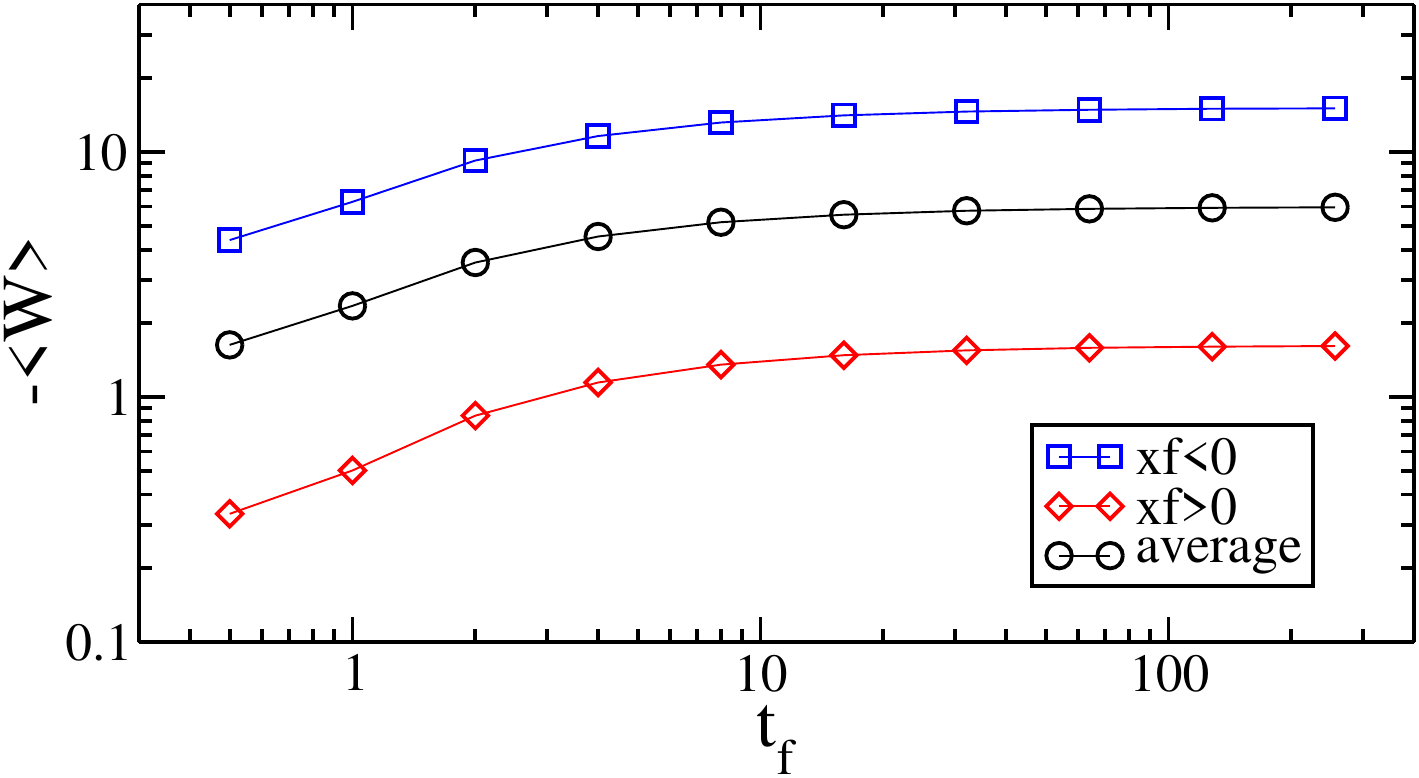}
  \caption{\label{fig3} Average useful work, $-\left<W\right>$, as a function of length of the
    extraction process, $t_f$. Squares: useful work extracted for $xf$ being negative at the initial time,
    diamonds: useful work for $xf$ being initially positive, circles: useful work averaged over the two
    measurement outcomes. 
    Initial and final stiffness $k=1$, persistence time $\tau_p=0.4$, self-propulsion strength $a=10$, 
    temperature $T=1$ and friction constant $\gamma=1$.}
\end{figure}

We find that machine learned stiffness
protocols allow us to extract significantly more work than the stepwise changes
of the stiffness, see Fig. \ref{fig1}b for work extracted in the limit of very long processes.
Furthermore, we find that the largest amount of work is extracted for persistence time of about
$\tau_p\approx 0.2$, which is shorter than that suggested by the stepwise change of the stiffness. 
For finite-length processes the learned protocols exhibit discontinuous changes
of the stiffness at both initial, $t=0$, and final, $t=t_f$, times, see Fig. \ref{fig2}.
The changes at the initial time are opposite to the expectation, \textit{e.g.} for product $xf$ negative at
$t=0$ the stiffness constant first decreases discontinuously and then increases above its initial value.
Similar behavior was found by Garcia-Millan \textit{et al.} \cite{GarciaMillan2025}.
Finally, as shown in Fig. \ref{fig3}, the work extracted increases with increasing duration of the process
and saturates in the large extraction time limit.

\textit{Discussion} --- We showed that non-trivial correlations between self-propulsion and position
that are present in systems of active particles can be exploited to devise a Szilard engine.
The work extracted from such an engine depends significantly on the protocol used for work
extraction. Simple stepwise protocols can demonstrate the feasibility of the engine but
machine learned protocols are able to extract significantly more work. The learned
protocols involve rather large stiffness changes and thus are outside of the linear response regime.
They involve discontinuous jumps at both initial and final times of the extraction process.
The jumps at the initial time are in the direction opposite to the one expected. This finding
combined with earlier results of Garcia-Millan \textit{et al.} \cite{GarciaMillan2025} suggests
that such initial jumps are a general feature of optimal protocols. 

While we explicitly analyzed work extraction in a single process, our approach can be used to set up
a simple cyclic active information engine. The engine would start with the active particle in its
stationary state. Then, the direction of the particle's self-propulsion relative to the direction of the confining
force wold be measured and the time-dependent protocol would be applied. The cycle would end with
the particle relaxing back to its stationary state with the stiffness constant equal to that at the
beginning of the cycle. This simple cyclic engine scheme can potentially be significantly
improved by using learned protocols for all parts of the cycle \cite{Frim2022,Chatterjee2025}.

Finally, while we have analyzed the very simple AOUP model in one spatial dimension, our
design can be easily adapted to the ABP model in two spatial dimensions. In this
case one would have to simulate ABP trajectories in order to evaluate the time-dependent average
second moment of the position, making finding optimal protocols more computationally time-consuming.
On the other hand, since the ABP model is a realistic model for active colloidal particles
\cite{Howse2007,Palacci2010}, the resulting Szilard engine protocols could then be tested experimentally. 

\textit{Acknowledgments} --- I gratefully acknowledge the support of NSF Grant No.~CHE 2154241.

\textit{Data availability statement} --- 
The data that support the findings of this article are openly available \cite{data}.

\textit{Appendix} --- Here we provide some additional details about the machine learning procedure.
The code consists of a Fortran 90 program that initializes 50 demons, 
finds the five best performers and restarts the demons, and a Fortran 77 program that
mutates a specific instance of the network and calculates the resulting work.

To determine the work for a given instance, we solve three coupled ordinary differential equations for
the second moments using a fourth order Runge-Kutta routine \cite{NumericalRecipes} with time step of
$10^{-4}$. The instances of the network are initialized using independent Gaussian random numbers with
variance $\sigma^2=0.01$. The first 1000 ``mutations'' involve adding independent Gaussian random numbers
with the same variance to the network parameters.
This is followed by 1000 ``mutations'' with variance $4\times 10^{-4}$, 
2000 ``mutations'' with variance $2.5\times 10^{-5}$ and  6000 ``mutations'' with variance $10^{-6}$.
We checked that our training procedure, when applied to a single passive particle,
recovers the analytical protocols derived by Schmiedl and Seifert \cite{Schmiedl2007}.

We confirmed that the useful work and learned protocols are not very sensitive to changes in the mutation
variances. We also examined the dependence of our results on the network architecture. 
We found that reducing the network width significantly changes the resulting work values and protocols, but
decreasing the depth or increasing the size has a very small impact.

\end{document}